\documentclass[reprint,english,aps,twocolumn,prx,superscriptaddress,preprintnumbers,amsmath,amssymb]{revtex4-1}

\usepackage{amsmath,amssymb}
\usepackage{graphicx}
\usepackage{rotating}
\usepackage{dsfont}
\usepackage{color}
\usepackage[dvips]{epsfig}     
\usepackage{multirow}
\definecolor{refcol}{RGB}{178,34,34}
\usepackage{microtype}
\usepackage[colorlinks,linkcolor=refcol,citecolor=refcol,urlcolor=refcol]{hyperref}
\usepackage{breakurl}
\usepackage{charter} 
\usepackage[charter]{mathdesign}

\usepackage[utf8]{inputenc}

\graphicspath{{./figures/}}

\def\eq#1{(\ref{#1})}
\def\Eq#1{Eq.~(\ref{#1})}
\def\Fig#1{Fig.~\ref{#1}}

\def\s0#1#2{\mbox{\small{$ \frac{#1}{#2} $}}}
\def\0#1#2{\frac{#1}{#2}}

\def\eq#1{(\ref{#1})}

\def\Eq#1{Eq.~(\ref{#1})}
\def\Fig#1{Fig.~\ref{#1}}

\definecolor{blue}{rgb}{0,0,1}

\definecolor{green}{rgb}{0,1,0}

\definecolor{red}{rgb}{1,0,0}

\definecolor{fabcol}{rgb}{0.56,0.00,1.00}

\newcommand{\tr}{\mathrm{tr}}

\newcommand{\be}{\begin{eqnarray}}
\newcommand{\ee}{\end{eqnarray}}

\usepackage{babel}
\makeatother

\begin{document}

\title{Strangeness neutrality and baryon-strangeness correlations}

\author{Wei-jie Fu}
\affiliation{School of Physics , Dalian University of Technology, Dalian, 116024, P.R. China}

\author{Jan M. Pawlowski}
\affiliation{Institut
 f\"{u}r Theoretische Physik, Universit\"{a}t Heidelberg,
 Philosophenweg 16, 69120 Heidelberg, Germany}
 
 \author{Fabian Rennecke}
\email[E-mail: ]{frennecke@bnl.gov\\[-1ex]}
\affiliation{Physics Department, Brookhaven National Laboratory,
  Upton, NY 11973, USA}

\begin{abstract}
  We derive a simple relation between strangeness neutrality and
  baryon-strangeness correlations. In heavy-ion collisions, the former
  is a consequence of quark number conservation of the strong
  interactions while the latter are sensitive probes of the character
  of QCD matter. This relation allows us to directly extract
  baryon-strangeness correlations from the strangeness chemical
  potential at strangeness neutrality. The explicit calculations are
  performed within a low energy theory of QCD with 2+1 dynamical quark
  flavors at finite temperature and density.  Non-perturbative quark and hadron
  fluctuations are taken into account within the functional
  renormalization group.
 The results show the pronounced sensitivity of baryon-strangeness
  correlations on the QCD phase transition and the crucial role that
  strangeness neutrality plays for this observable. 

\end{abstract}

\maketitle

\textbf{Introduction.}
Extracting the phase diagram of QCD as a function of temperature $T$
and baryon chemical potential $\mu_B$ from hadronic final states is a
main goal
but also a main challenge 
in ultra-relativistic heavy-ion
collisions. A detailed understanding of
QCD matter in the hot and dense medium is indispensable for the
interpretation and prediction of experimental data. The situation is
further complicated by the fact that under the conditions of a
heavy-ion collision, the phase diagram is not only spanned by $T$ and
$\mu_B$, but for instance also other chemical potentials,
electromagnetic fields and various timescales. This is
relevant for beam-energy scan experiments aiming at exploring the QCD
phase diagram \cite{Gazdzicki:995681, *BESwp, *Friman:2011zz,
  *Kekelidze:2016hhw, *GALATYUK201441, *Sako:2014fha}.  Fortunately,
conservation laws can help to constrain some of these parameters.

The hadrons reaching the detectors in heavy-ion experiments inherit
the properties of the QCD medium at freeze-out. Since the typical
freeze-out time is many orders of magnitude shorter than the timescale
of flavor-changing weak interactions, the net quark number
conservation of the strong interactions has to be taken into
account. Consequently, there is a chemical potential associated to the
conserved quark number of each quark flavor. Focusing on the three
lightest flavors, up $u$, down $d$ and strange $s$, the associated
chemical potentials are conventionally given by linear combinations of
baryon, charge and strangeness chemical potentials $\mu_B$, $\mu_Q$
and $\mu_S$. With the chemical potential flavor-matrix
\begin{align}\label{eq:mumat}
  \mu = \text{diag}\bigg(\!\frac{1}{3}\mu_B+\frac{2}{3}\mu_Q,
  \frac{1}{3}\mu_B-\frac{1}{3}\mu_Q, \frac{1}{3}\mu_B-\frac{1}{3}\mu_Q
  - \mu_S\!\!  \bigg)\,,
\end{align}
their coupling to the quarks, $q = (u,d,s)^T$, is given by
$\bar q \gamma_0 \mu\, q$. In this work, we want to focus on
strangeness and baryon number and therefore assume $\mu_Q = 0$ for the
sake of simplicity. This corresponds to isospin symmetric
matter. Since the incident nuclei do not carry strangeness, the
net-strangeness
$\langle S \rangle \sim \langle\bar s \gamma_0 s\rangle$ is fixed from
the initial conditions of the collision. The condition
$\langle S \rangle = 0$ is called strangeness neutrality. Due to the
peculiar beam-energy dependence of the net-baryon rapidity spectrum,
the net-baryon number density
$\langle B \rangle \sim \langle\bar q \gamma_0 q\rangle$ in the
quark-gluon plasma (QGP) at central rapidities depends on the beam
energy, see, e.g., \cite{Adamczyk:2017iwn}.  We therefore work with
the standard assumption that $\mu_B$ is a parameter we may choose
freely, while $\mu_S$ is fixed through quark number conservation,

Strangeness is particularly interesting since strange particles are
only created by collisions in the first place. This makes them
valuable probes of the matter created in heavy-ion collisions
\cite{Koch:1986ud}. In a recent work \cite{Fu:2018qsk} we have
investigated the effect of imposing strangeness neutrality on the
phase structure and thermodynamics of QCD. Strangeness neutrality has
a sizable impact on the QCD equation of state and the phase diagram,
owing to an intricate interplay of strangeness coupled to meson-,
baryon- and quark dynamics at finite $T$ and $\mu_B$, Furthermore,
this interplay also leads to the observation that a finite $\mu_S$ is
required to enforce $\langle S \rangle = 0$. Hence, we are led to the implicitly
defined function
\begin{align}\label{eq:mus0}
\mu_{S0}(T,\mu_B) = \mu_S(T,\mu_B)\big|_{\langle S \rangle = 0}\,.
\end{align}
Since $\mu_B$ couples to all quark flavors equally, increasing $\mu_B$ also increases the number of strange
quarks over anti-strange quarks. To ensure strangeness neutrality, a
finite $\mu_S$ is necessary for compensation; see also, e.g.,
\cite{Letessier:1993hi}. In the hadronic phase the dominant degrees of
freedom are either open strange mesons or strange baryons, depending
on $\mu_B$. But while both couple to $\mu_S$, only the latter couple
to $\mu_B$. Thus, it is intuitively clear that $\mu_{S0}$ is a
non-trivial function that is intimately tied to the nature of QCD
matter

In this letter we will demonstrate that this deep connection between
strangeness neutrality, which is a consequence of quark number
conservation, and the dynamical interplay of hadrons and quarks, which
is interweaved with the phase structure of QCD, can be made
explicit. This is achieved by establishing an exact relation between
$\mu_{S0}$ and the baryon-strangeness correlation $C_{BS}$, see
\eq{eq:cbs}.  This correlation has been introduced in
\cite{Koch:2005vg} as a sensitive probe of the nature of QCD
matter. We will exploit said relation to compute $C_{BS}$ at various
$T$ and $\mu_B$ and carve out the important role that strangeness
neutrality plays for this quantity. To this end, we employ a Polyakov
loop enhanced quark-meson model (PQM) with 2+1 dynamical quark flavors
as a low-energy effective theory of QCD. Non-perturbative quantum,
thermal and density fluctuations are taken into account with the
functional renormalization group.
Within this approach, quark-meson models are naturally embedded in
QCD \cite{Mitter:2014wpa, *Braun:2014ata, *Rennecke:2015eba, *Cyrol:2017ewj}.


\vspace{1ex}\textbf{Strangeness neutrality \& baryon-strangeness correlations.}
Generalized susceptibilities of conserved charges play a central role
for theoretical and experimental studies of the QCD phase
structure. This is due to the fact that the closely related cumulants
of particle number distributions are directly sensitive to the growing
correlation length at the phase transition \cite{Stephanov:2008qz}. In
the present context at $\mu_Q = 0$, the generalized susceptibilities
are defined as chemical
potential derivatives of the pressure $p$,
\begin{align}\label{eq:sus}
  \chi_{ij}^{BS}(T,\mu_B,\mu_S) = \frac{\partial^{i+j}
  p(T,\mu_B,\mu_S)/
  T^4}{\partial\hat\mu_B^i\partial\hat\mu_S^j}\,,
\end{align}
with $\hat\mu = \mu/T$. Baryon number and strangeness are then given
by
\begin{align}\nonumber
  \langle B \rangle &= \langle N_B -N_{\bar B} \rangle =
                      \chi_{1}^{B}\, VT^3\,,\\
  \langle S \rangle &= \langle N_{\bar S} - N_S \rangle =
                      \chi_{1}^{S}\, VT^3\,,
\label{eq:pnums}
\end{align}
where $V$ is the spatial volume. We drop the superscript of the
susceptibilities if the corresponding subscript is zero. The
baryon-strangeness correlation $C_{BS}$ \cite{Koch:2005vg} relevant
for the present work reads at strangeness neutrality
\begin{align}\label{eq:cbs}
  C_{BS}(T,\mu_B,\mu_{S0}) =
  -3 \frac{\langle BS \rangle}{\langle S^2 \rangle}=  -3 \frac{\chi_{11}^{BS}(T,
  \mu_B,\mu_{S0})}{\chi_{2}^{S}(T,\mu_B,\mu_{S0})}\,.
\end{align}
Since the pressure is a function of $T$, $\mu_B$ and $\mu_S$, requiring
strangeness neutrality implicitly fixes one of these variables as
function of the others. Here, we choose the strangeness chemical
potential, giving rise to \Eq{eq:mus0}. $\langle S \rangle = 0$
implies
\begin{align}\nonumber
0 &= \frac{d}{d\hat\mu_B} \chi_{1}^{S}(T,\mu_B,\mu_{S0})\\
 & = \chi_{11}^{BS}(T,\mu_B,\mu_{S0}) +\chi_{2}^{S}(T,\mu_B,\mu_{S0}) 
\frac{\partial \hat\mu_{S0}}{\partial\hat\mu_B}\,.
\label{eq:strneutr1}
\end{align} 
This simple equation implicitly defines $\mu_{S0}$ and we arrive at
our central result,
\begin{align}\label{eq:cbs1}
  \frac{\partial \mu_{S0}(T,\mu_B)}{\partial\mu_B} = \frac{1}{3}
  C_{BS}(T,\mu_B,\mu_{S0})\,.
\end{align}
Hence, together with the initial condition $\mu_{S0}(\mu_B = 0) = 0$,
$\mu_{S0}$ can be extracted directly from $C_{BS}$ at strangeness
neutrality. By integrating the experimentally measured $C_{BS}$ over
the beam energy, one can extract the strangeness chemical potential at
the freeze-out for isospin symmetric matter. Conversely,
$C_{BS}$ is given by the slope of $\mu_{S0}(\mu_B)$. Most strikingly,
$C_{BS}$ has been introduced in \cite{Koch:2005vg} as a diagnostic for
the nature of QCD matter. As argued in this work, this can be
understood by explicitly examining $C_{BS}$ at strangeness neutrality, 
see \eq{eq:cbs}. Assuming that the system is deep in the deconfined regime, all
strangeness is carried by $s$ and $\bar s$ and there is a strict
relation between the baryon number carried by strange particles,
$B_s$, and strangeness, $B_s = S/3$. Furthermore, due to asymptotic
freedom, there is no correlation between different quark flavors in
this regime; it is 
a system of dilute current quarks. \Eq{eq:cbs} then
implies $C_{BS} = 1$.

The situation is drastically different in the confined phase. Baryons
can carry both baryon number and strangeness, while mesons can only
carry strangeness. Thus, the denominator in \Eq{eq:cbs},
$\chi_{2}^{S}$, includes both open strange mesons and baryons, while
the numerator, $\chi_{11}^{BS}$, only includes strange baryons. Hence,
one generally finds $C_{BS}\neq 1$ in the confined hadronic phase. For
$C_{BS} < 1$ the system is dominated by the fluctuations of open
strange mesons and for $C_{BS} > 1$ it is dominated by strange
baryons. Since cumulants of net particle numbers are experimental
accessible, $C_{BS}$ indeed serves as a sensitive probe of the
composition of QCD matter, in particular of its strangeness content
\cite{Bazavov:2013dta, Bazavov:2014xya}. It is remarkable that
$C_{BS}$ can be directly related to $\mu_{S0}$ via strangeness
neutrality with \Eq{eq:cbs1} for any $T$ and $\mu_B$. This establishes
a direct connection between quark number conservation and the phases
of QCD.

We note that the situation becomes a little more complicated at
$\mu_Q \neq 0$. In this case \Eq{eq:sus} trivially generalizes to
$\chi_{ijk}^{BSQ}$. In addition to $\chi_{1}^{S}= 0$, quark number
conservation also implies, for instance,
$\chi_{1}^{Q}/\chi_{1}^{B} = r$, where $r$ is a constant. This also
implicitly defines the function $\mu_{Q0}(T,\mu_B)$ and leads to a
generalized form of \Eq{eq:cbs1},
\begin{subequations}\label{eq:mus0full}
\begin{align}
  \frac{\partial \mu_{S0}}{\partial \mu_B} &= 
                                             \frac{1}{3} C_{BS}
                                             -\frac{\chi_{11}^{QS}}{
                                             \chi_{2}^{S}}\frac{\partial\mu_{Q0}}{\partial\mu_B}\,,
\end{align}
with
\begin{align}
  \frac{\partial \mu_{Q0}}{\partial \mu_B} &=
                                             \frac{\chi_{11}^{BS}(\chi_{11}^{SQ}-r
                                             \chi_{11}^{BS})-\chi_{2}^{S}(\chi_{11}^{BQ}-r
                                             \chi_{2}^{B})}{\chi_{2}^{S}(\chi_{2}^{Q}-r\chi_{11}^{BQ})
                                             -\chi_{11}^{SQ}(\chi_{11}^{SQ}-r
                                             \chi_{11}^{BS})}\,.
\end{align}
\end{subequations}
The dependence on $T$ and $\mu_B$ of all quantities above is
implied. The right hand side of these equations is given by ratios of
different baryon, strangeness and charge correlations and therefore
can be interpreted in an analogous manner to the discussion above,
cf.\ \cite{Majumder:2006nq}. Furthermore, since the involved
susceptibilities can be measured, our main conclusion is not altered
for the isospin-asymmetric case. It is worth noting that
\Eq{eq:mus0full} generalizes the relations used on the lattice to
implement the freeze-out conditions in an expansion about
$\mu_B/T = 0$ to any $T$ and $\mu_B$ \cite{Bazavov:2012vg,
  *Borsanyi:2013hza, *Bazavov:2017tot}.


\vspace{1ex}\textbf{Low-energy effective theory \& fluctuations.}
In the following, we compute baryon-strangeness correlations with the
help of \Eq{eq:cbs1}. The impact of strangeness neutrality at
  finite baryon chemical potential is studied within a low-energy
  effective theory of QCD as initiated in \cite{Fu:2018qsk}. In order
to capture the main features of strangeness, quantum, thermal and
density fluctuations of open strange mesons, strange baryons and
quarks have to be taken into account. Since kaons are pseudo-Goldstone
bosons of spontaneous chiral symmetry breaking, they are the most
relevant strange degrees of freedom in the mesonic sector. Moreover,
chiral symmetry dictates that if kaons are included as effective
low-energy degrees of freedom, all other mesons in the lowest scalar
and pseudoscalar meson nonets have to be included as well. By coupling
quarks to a uniform temporal gluon background field
$\bar A_0 = \bar A_0^{(3)}\, t^3 + \bar A_0^{(8)}\, t^8$, with
$t^c\in SU(3)$, (statistical) confinement is taken into account. Below
the deconfinement transition temperature $T_d$ predominantly
three-(anti-)quark states contribute and baryons, instead of quarks,
are the prevailing fermionic degrees of freedom below
$T_d$.
In total, this gives rise to a 2+1 flavor PQM with the Euclidean
effective action ($\beta = 1/T$)
\begin{align}\nonumber
  \Gamma_k &= \int_0^{\beta} \!\!dx_0 \int \!\!d^3x\Big\{
             \bar q \big(\gamma_\nu D_\nu+\gamma_\nu C_\nu\big)q + h\,
             \bar q\,\Sigma_5 q\\
           &\quad+\text{tr}\big(\bar D_\nu\Sigma\!\cdot\!\bar D_\nu
             \Sigma^\dagger\big)+\widetilde U_k(\Sigma) + U_\text{glue}(L,\bar L)
             \Big\}\,.
\label{eq:ea}          
\end{align}
Quantum, thermal and density fluctuations of modes with Euclidean
momenta $k \leq |p| \lesssim 1$ GeV have been integrated out. The
gauge covariant derivative is
$D_\nu = \partial_\nu - i g \delta_{\nu 0} \bar A_0$. The scalar and
pseudoscalar mesons are encoded in the flavor matrix
$\Sigma = T^a (\sigma_a + i \pi_a)$, where the $T^a$ generate
$U(N_f)$, and $\Sigma_5 = T^a (\sigma_a+i \gamma_5\pi_a)$, see
e.g.\ \cite{Rennecke:2016tkm}.  The couplings of quarks and
mesons to the chemical potential $\mu$ in \Eq{eq:mumat} is achieved by
formally introducing the vector source $C_\nu = \delta_{\nu 0} \mu$
and defining the covariant derivative acting on the meson fields
$\bar D_\nu \Sigma = \partial_\nu \Sigma +[C_\nu,\Sigma]$,
\cite{Kogut:2001id}.

Spontaneous chiral symmetry breaking is captured by the meson
effective potential $\tilde U_k(\Sigma)$, which consists of a fully
$U(N_f)_L\times U(N_f)_R$ symmetric part plus pieces that explicitly
break chiral symmetry through finite current quark masses and
$U(1)_A$ through the axial anomaly. 
The deconfinement phase transition is captured statistically by
including an effective potential for the gluon background field
$U_\text{glue}(L,\bar L)$, formulated in
terms of the order parameter fields for deconfinement, the Polyakov
loops $L = \tr_c \exp(i g \bar A_0 / T)$,
$\bar L = \tr_c [\exp(i g \bar A_0 / T)]^\dagger$. The strategy of Polyakov loop enhanced
effective models is to use a potential that is fitted to the lattice
equation of state of Yang-Mills theory and to include the effects of
dynamical quarks through the coupling to the gluonic background
in the quark covariant derivative. For a recent review see \cite{Fukushima:2017csk}. 
We use the parametrization of the Polyakov loop potential 
put forward in \cite{Lo:2013hla}, since it
captures the lowest-order Polyakov loop susceptibilities which
directly contribute to the particle number susceptibilities
\cite{Fu:2016tey}.

\begin{figure}[t]
\centering
\includegraphics[width=1.02\columnwidth]{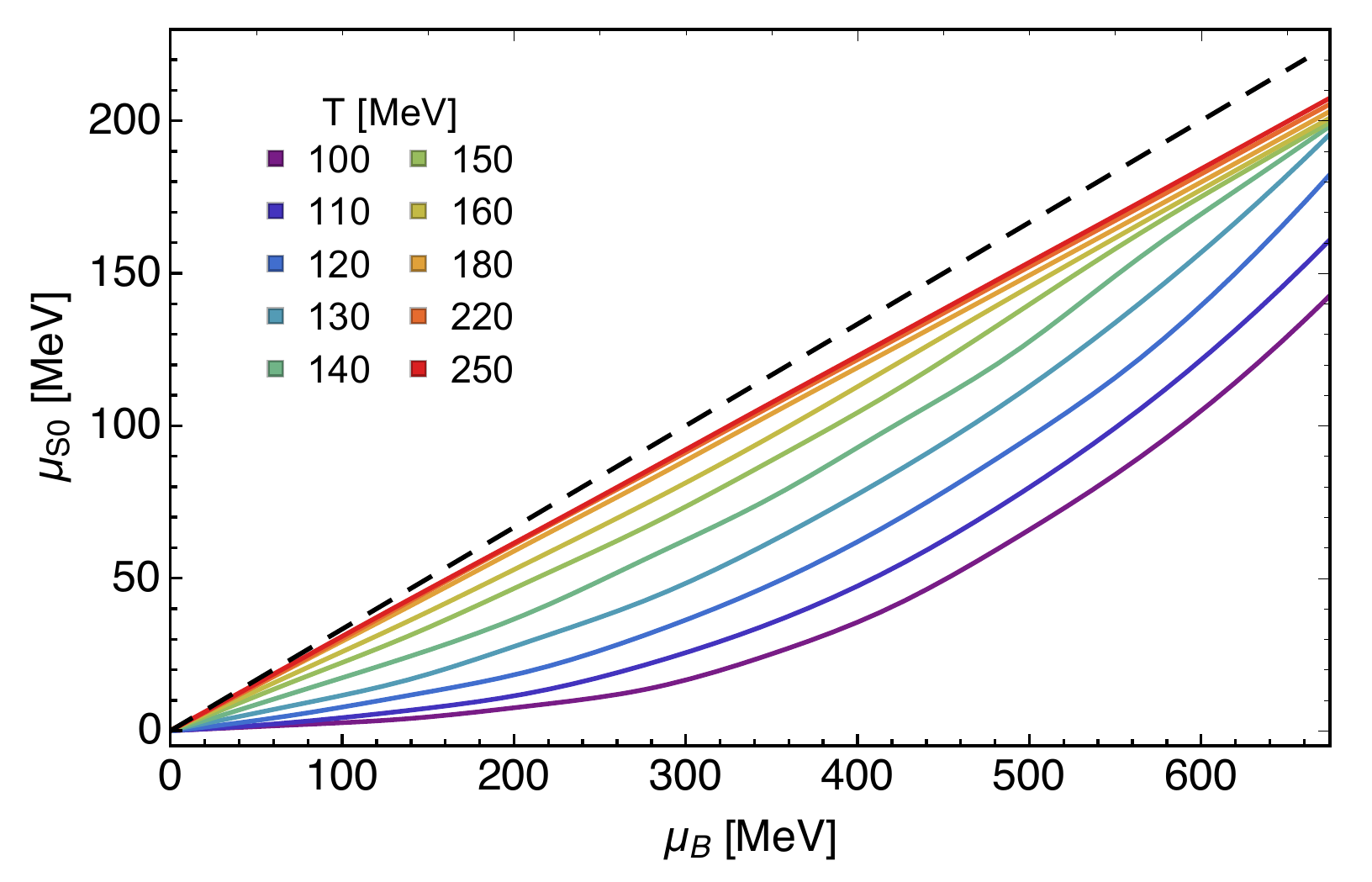}
\caption{Strangeness chemical potential at strangeness neutrality,
  $\mu_{S0}$, as a function of the baryon chemical potential $\mu_B$
  for various temperatures $T$ (solid lines). T is increasing from
  bottom to top from 100 MeV to 250 MeV. The dashed line corresponds
  to the asymptotic limit of free quarks. In this limit one finds
  $C_{BS} = 1$ which leads to $\mu_{S0} = \mu_B / 3$ according to
  \Eq{eq:cbs1}.}
\label{fig:musofmub}
\end{figure}

Owing to the intricate interplay of meson, baryon and quark dynamics
that contribute to strangeness neutrality and, as a consequence of
\Eq{eq:cbs1}, also to $C_{BS}$, accounting for fluctuations of these
degrees of freedom is indispensable. In \cite{Fu:2018qsk} we
demonstrated that open strange meson fluctuations are crucial for
strangeness neutrality, exacting a treatment beyond mean-field. Here,
this is achieved by solving the renormalization group flow equation
for the effective action $\Gamma_k$ using the functional
renormalization group (FRG) \cite{Wetterich:1992yh},
\begin{align}\label{eq:fleq}
\partial_t \Gamma_k = \frac{1}{2} \text{Tr}\, \Big[ \big(\Gamma_k^{(2)}[\Phi] + R_k\big)^{-1}\partial_t R_k\Big]\,,
\end{align}
with $t = \ln (k/\Lambda)$. $\Gamma_k^{(2)}[\Phi]$ is the matrix of
second functional derivatives of the effective action with
respect to the fields $\Phi = (q,\bar q, \Sigma, L, \bar L)$. $R_k$
implements infrared-regularization at momenta $p^2 \approx k^2$ and
the trace involves the integration over loop-momenta, the color-,
flavor- and spinor-traces as well as the sum over different particle
species. Solving \Eq{eq:fleq}
amounts to successively integrating out fluctuations starting from the
initial action $\Gamma_{k=\Lambda}$, with $\Lambda = 900$ MeV in our
case, down to the full quantum effective action $\Gamma_{k=0}$. The
FRG provides a non-perturbative regularization and renormalization
scheme for the resummation of an infinite class of Feynman diagrams;
see, e.g., \cite{Berges:2000ew, *Pawlowski:2005xe, *Gies:2006wv, *Schaefer:2006sr,*Braun:2011pp} for
QCD-related reviews.

It was shown in \cite{Fu:2018qsk} that this approach leads to a very
good agreement with the results of lattice QCD for the equation of
state at vanishing $\mu_B$ and finite $\mu_B/T$ within the region
accessible on the lattice. For further technical details on the model
and on the RG flow equations we refer to this work. Since the main
qualitative features relevant for strangeness dynamics are captured by
this approach, we will use it to compute the strangeness chemical
potential at strangeness neutrality, $\mu_{S0}(T,\mu_B)$, and then use
\Eq{eq:cbs1} to extract the baryon-strangeness correlation $C_{BS}$.


\vspace{1ex}\textbf{Numerical results.}
From the solution of the RG flow equation \eq{eq:fleq} we obtain the
full quantum effective action $\Gamma_0[\Phi]$. With the solution of
the quantum equation of motion
$(\delta \Gamma_0[\Phi]/\delta\Phi)|_{\Phi = \Phi_\text{EoM}}= 0$, the
pressure is given by $p = -\Gamma_0[\Phi_\text{EoM}]/\beta V$. With
this we are in the position to extract the generalized
susceptibilities according to \Eq{eq:sus}. First, we compute the
strangeness number $\langle S \rangle$ in \Eq{eq:pnums} as a function
of $T$, $\mu_B$ and $\mu_S$. It turns out out that for any $T$ and
$\mu_B$ it is always possible to find a $\mu_S = \mu_{S0}$ such that
$\langle S \rangle (T,\mu_B,\mu_S = \mu_{S0}) = 0$. We restrict
ourselves to $\mu_B \leq 675$ MeV since our model fails to capture
important qualitative features of the theory at larger $\mu_B$, cf.\
\cite{Fu:2018qsk}. The transition is a crossover in this range. The resulting $\mu_{S0}$ is shown in
\Fig{fig:musofmub}. The characteristic shape of $\mu_{S0}(\mu_B)$ can
be understood qualitatively from our discussion of \Eq{eq:cbs1}. At
small $\mu_B$ strangeness is dominated by open strange mesons, so
$C_{BS} < 1$ and hence $\mu_{S0}(\mu_B)$ has a slope smaller than
$1/3$. At larger $\mu_B$ strange baryons become dominant resulting in
a slope larger than $1/3$. At large temperatures the system undergoes
a crossover to the deconfined phase with a pseudocritical temperature
of $T_d \approx 155$ MeV in our model. Asymptotically,
$\mu_{S0}(\mu_B)$ approaches the dashed black line. However, since the
Polyakov loops are still smaller than one at $T = 250$ MeV in our
computations, implying that the system is not fully deconfined, this
asymptotic limit is not fully reached here. Still $\mu_{S0}(\mu_B)$ is
approximately linear already at $T \approx 180$ MeV, with a slope only
slightly smaller than $1/3$.

\begin{figure}[t]
\centering
\includegraphics[width=1.02\columnwidth]{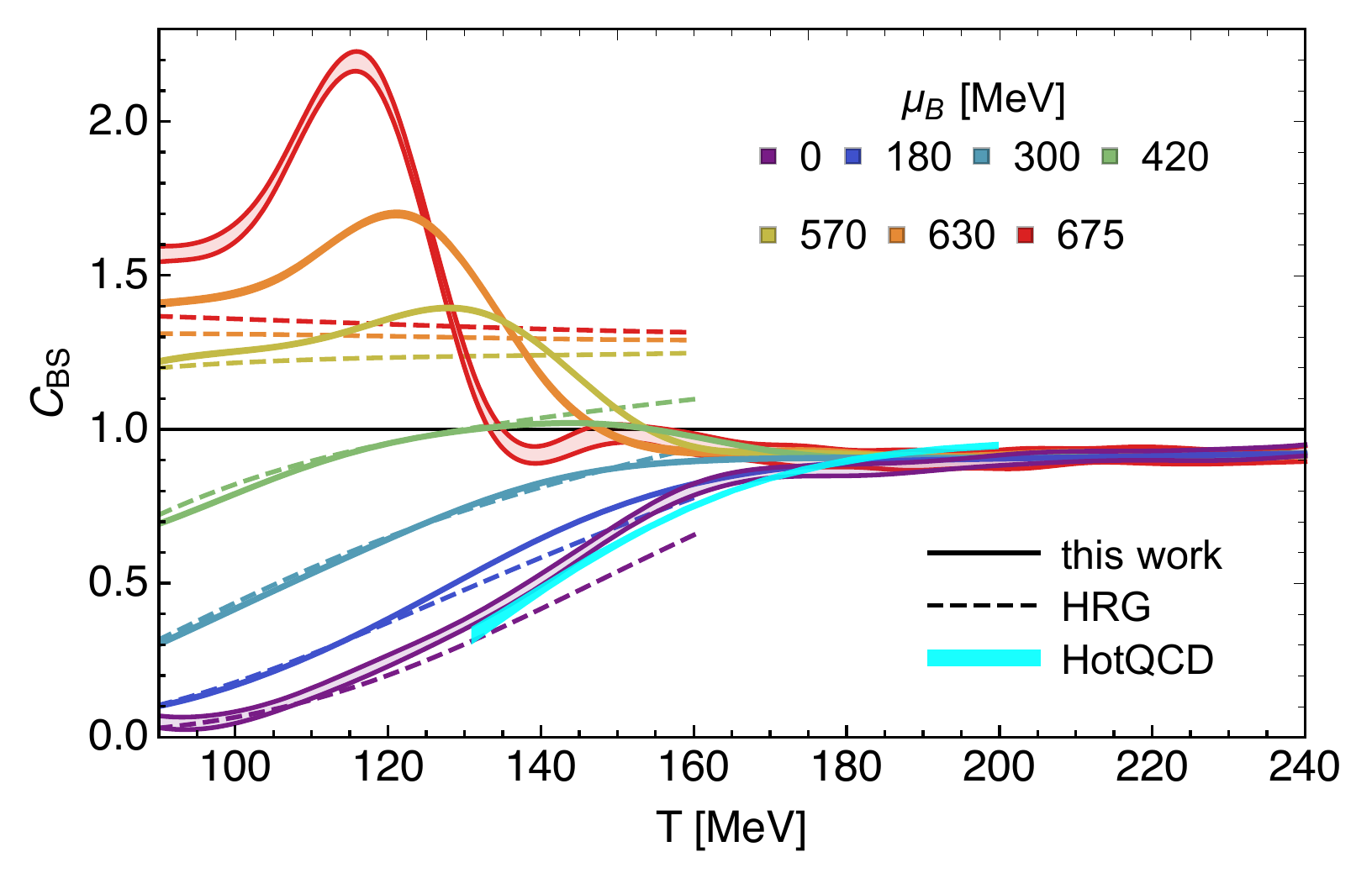}
\caption{Baryon-strangeness correlation $C_{BS}$ as a function of
  temperature $T$ for different baryon chemical potential $\mu_B$ at
  strangeness neutrality. $\mu_B$ increases from 0 to 675 MeV from
  bottom to top. We compare our results to the hadron resonance gas (HRG) containing only experimentally observed resonances \cite{BraunMunzinger:2003zd, *Tanabashi:2018oca}. At $\mu_B = 0$ we also compare to the result of lattice
  QCD \cite{Bazavov:2013dta, Bazavov:2014xya, Bazavov:2017dus, *Bazavov:2018mes, *swagato}. The thin black line indicates the free
  quark limit . The errors reflect the 95\% confidence
  level of a cubic spline interpolation of our numerical data.}
\label{fig:cbsoft}
\end{figure}

We extract the
baryon-strangeness correlations $C_{BS}$ from our result in
\Fig{fig:musofmub} via \Eq{eq:cbs1}. This is shown
in \Fig{fig:cbsoft}. We find good agreement with the results of lattice QCD at vanishing $\mu_B$, highlighting that we capture the relevant effects quite accurately.
Following our discussion above, we see that with increasing $\mu_B$ in the hadronic phase the baryon-strangeness correlations change from being dominated by
the dynamics of open strange mesons to being dominated by strange baryons. 
At $\mu_B \!\gtrsim\! 420$ MeV $C_{BS}(T)$ develops a
non-monotonicity. It first grows with $T$
since with increasing temperature more strange baryons
can be excited and a larger $\mu_S$ has to be chosen
in order to enforce strangeness neutrality. The resulting increasing
slope of $\mu_{S0}$ then directly translates to a rising $C_{BS}$
through \Eq{eq:cbs1}. For temperatures above the pseudocritical
transition temperature, quark dynamics eventually take over, driving
the system towards its asymptotic limit. 

As a result of this dynamical interplay, $C_{BS}(T)$ shows non-monotonous behavior and develops a pronounced maximum already at moderate $\mu_B$. The sharper the crossover, the stronger this effect becomes. The maximum is located exactly in the crossover region. We therefore find a distinct sensitivity of baryon-strangeness
correlations to the chiral phase transition at finite $\mu_B$. This is
potentially relevant for experimental measurements of $C_{BS}$: If the
freeze-out is close to the phase transition, we predict a steep rise
of $C_{BS}$ with decreasing beam-energy.

\begin{figure}[t]
\centering
\includegraphics[width=1.02\columnwidth]{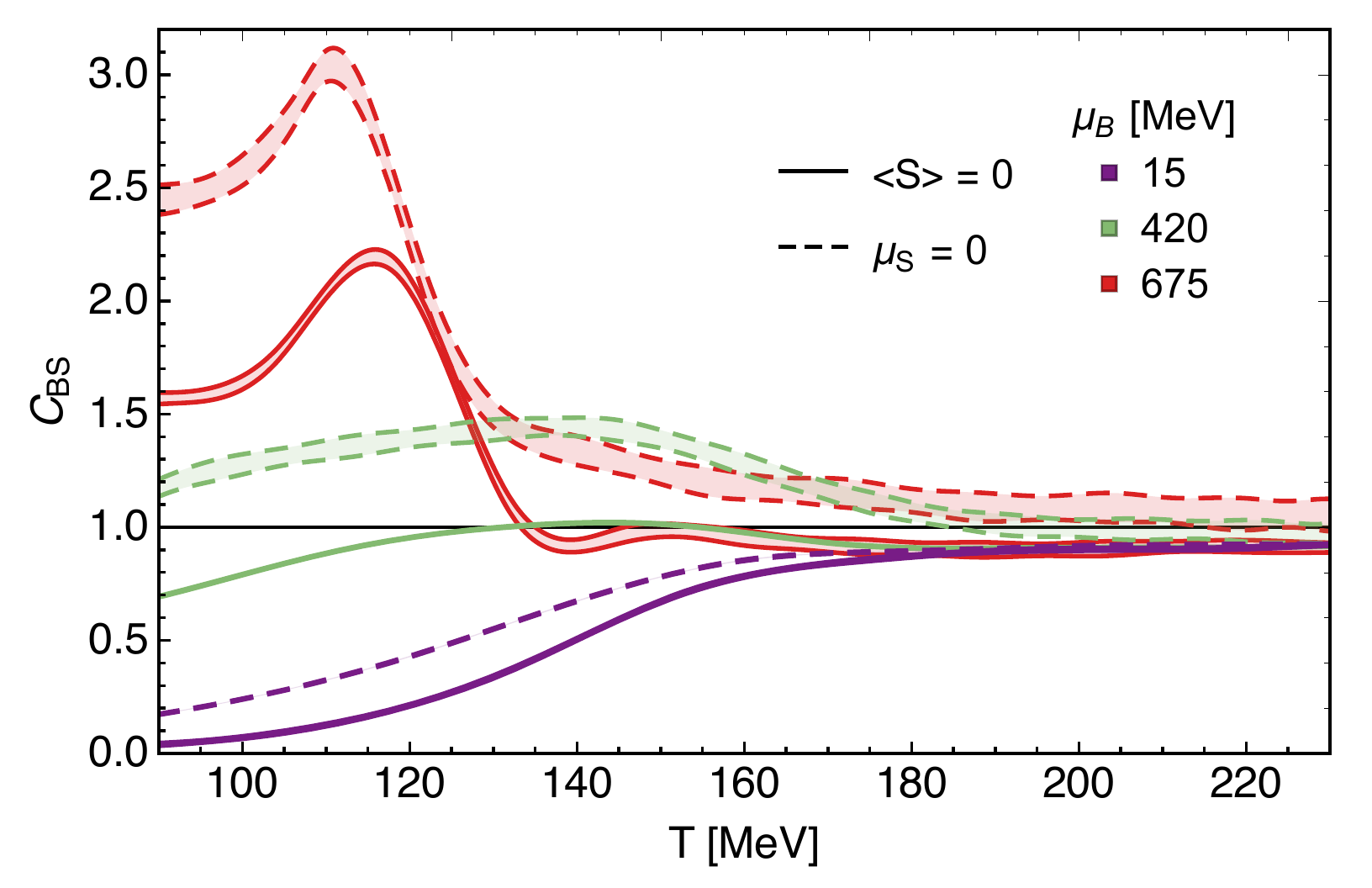}
\caption{Comparison between $C_{BS}$ computed at strangeness
  neutrality, i.e.\ $\mu_S = \mu_{S0}$ (solid lines), and at
  $\mu_S = 0$ (dashed lines), where strangeness conservation is violated at finite $\mu_B$.}
\label{fig:cbsoftcomp}
\end{figure}

As seen in \Fig{fig:cbsoft}, the HRG predicts monotonous behavior of $C_{BS}(T)$ for any $\mu_B$ within the range of temperatures studied here. At large $\mu_B$ the non-monotonous behavior we find leads to a significant enhancement over the HRG predictions at temperatures around the chiral crossover transition. This could be related to long-range strangeness dynamics close to the phase transition. In general, the HRG is in good agreement with our results well below the phase transition temperature. At $\mu_B = 300$ MeV and 420 MeV we find good agreement also for larger temperatures. Whether this is physical or coincidental is not clear to us.
In \cite{Bazavov:2014xya} it was argued that the discrepancy between the lattice and the HRG at $\mu_B = 0$ could be due to yet undiscovered strange resonances which are not taken into account in the conventional HRG.
So perhaps the effects of missing open strange mesons are compensated by baryon fluctuations at intermediate densities. In any case, this leads us to another prediction for the measurement of $C_{BS}$: It should be significantly enhanced over the HRG prediction at small beam-energies 
\footnote{We start seeing a pronounced peak forming at $\mu_B \gtrsim 550$ MeV. If we take the scales of our computation at face value, this corresponds to beam energies of $\sqrt{s} \lesssim 5$ GeV. This would be beyond the reach of current beam-energy scan experiments, but within the range covered by future experiments, e.g.\ at FAIR, NICA or J-PARC.}. 
This would indicate a sharp crossover transition.

Finally, we want to explore the relevance of strangeness conservation on
baryon-strangeness correlations. To this end, we compute $C_{BS}$ also
at vanishing strangeness chemical potential, $\mu_S = 0$. Since there
is no connection between $\mu_S$ and $C_{BS}$ in this case, we have to
compute $\chi_{11}^{BS}$ and $\chi_{2}^{S}$ separately from the
pressure. The result in comparison to the one at strangeness
neutrality is shown in \Fig{fig:cbsoftcomp}. We find a sizable
enhancement of $C_{BS}$ if strangeness neutrality is not taken into
account, emphasizing the important role it plays here.

\vspace{1ex}\textbf{Summary.}
We have shown that there is an intimate 
relation between particle number conservation and the QCD
phase structure. This has been achieved by deriving a direct relation
between baryon-strangeness correlations and strangeness neutrality. We explicitly demonstrated the sensitivity of $C_{BS}$ on the phase structure and the relevance of strangeness neutrality for this
observable. 
The study of the critical behavior of $C_{BS}$ is deferred to future work. Since it is given by a ratio of second-order cumulants, it may not be sensitive to criticality at all. We emphasize that the sensitivity of $C_{BS}$ to the phase transition observed here is due to the change from hadronic to partonic dynamics, which is more pronounced at larger $\mu_B$ due to the specific dynamics that drive $C_{BS}$. A meaningful comparison between our theoretical prediction and experimental measurements may require, among other things, the description of net-kaons and protons instead of conserved charges, as well as non-equilibrium effects.

\vspace{1ex} \textit{Acknowledgments.} We thank Robert D. Pisarski
for discussions and Swagato Mukherjee for providing us with the lattice data for \Fig{fig:cbsoft}. F.R.\ is supported by the Deutsche
Forschungsgemeinschaft (DFG) through grant \mbox{RE 4174/1-1}. W.F.\ is
supported by the National Natural Science Foundation of China under
contract no.\ 11775041.  This work is supported by the ExtreMe Matter
Institute (EMMI). It is part of and supported by the DFG Collaborative
Research Centre "SFB 1225 (ISOQUANT)".


\bibliography{qcd-phase}

\end{document}